  \documentstyle[fleqn,mystyle,12pt]{article}   

  \newcommand{\beq}[1]{\begin{equation}\label{#1}}
  \newcommand{\eeq}{\end{equation}}             
  \newcommand{\bel}[1]{\begin{equation}\label{#1}}
  \newcommand{\eel}{\end{equation}}
  \newcommand{\belal}[1]{\begin{equation}\label{#1}\begin{array}{l}}
  \newcommand{\eelal}{\end{array}\end{equation}}
  \newcommand{\belac}[1]{\begin{equation}\label{#1}\begin{array}{c}}
  \newcommand{\eelac}{\end{array}\end{equation}}
  \newcommand{\R}{{\bf R}}

  \newcommand{\bP}{{\bf P}}
  \newcommand{\bS}{{\bf S}}

  \newcommand{\bX}{{\bf X}}

  \newcommand{\cB}{{\cal B}}

  \newcommand{\cH}{{\cal H}}

  \newcommand{\cK}{{\cal K}}
  \newcommand{\cL}{{\cal L}}

  \newcommand{\cR}{{\cal R}}
  \newcommand{\cS}{{\cal S}}

  \newcommand{\cX}{{\cal X}}

  \newcommand{\hA}{{\hat A}}
  \newcommand{\hB}{{\hat B}}

  \newcommand{\hH}{{\hat H}}

  \newcommand{\hM}{{\hat M}}
  \newcommand{\hN}{{\hat N}}

  \newcommand{\hU}{{\hat U}}

  \newcommand{\hX}{{\hat X}}
  \newcommand{\hY}{{\hat Y}}

  \newcommand{\rL}{{\rm L}}
  \newcommand{\rP}{{\rm P}}

  \newcommand{\al}{\alpha}                                             %
  \newcommand{\ep}{\epsilon}  

  \newcommand{\et}{\eta}                                               %
  \newcommand{\rh}{\rho}                                               %
  \newcommand{\si}{\sigma}                                             %
  \newcommand{\ta}{\tau}                                               %
  \newcommand{\ph}{\phi}                                               %

  \newcommand{\ps}{\psi}                                               %
  \newcommand{\om}{\omega}                                             %
  \newcommand{\De}{\Delta}                                             %

  \newcommand{\Tr}{\mbox{\rm Tr}}
  \newcommand{\Ex}{{\rm Ex}}
  
  \newcommand{\Var}{{\rm Var}}

  \newcommand{\dom}{\mbox{\rm dom}}
  \newcommand{\ran}{\mbox{\rm ran}}

  \newcommand{\iint}{\int\!\!\int}
  
  \newtheorem{Theorem}{Theorem}[section]
  
  \newtheorem{Lemma}[Theorem]{Lemma}
  \newtheorem{Corollary}[Theorem]{Corollary}

  \newenvironment{Proof}{\begin{trivlist}
    \item[\hskip \labelsep {\em \indent Proof.}]}{\qed\end{trivlist}}
  \newcommand{\qed}{{\em QED}}
  \newenvironment{Remark}{\begin{trivlist}
    \item[\hskip \labelsep {\bf \indent Remark.}]}{\end{trivlist}}
    
  \newcommand{\bold}[1]{\bf #1}
  
  \newcounter{bean}
  \newcommand{\hbX}{\hat{\bf X}}
  
  \title{\bf Quantum Limits of Measurements and \\
                    Uncertainty Principle}
  \author{                                                             %
  \sc        Masanao Ozawa\\ \\                                        %
  \small\em  Department of Mathematics, College of General Education\\
  \small\em  Nagoya University, Nagoya 464, Japan}
  \date{}

\begin{document}
  \itemsep=0in
  \maketitle
 

 \section{Introduction}                                               %
  \label{1}\setcounter{equation}{0}                                    %

In the theoretical considerations on quantum aspects of
optical communications, one of the most important programs is to 
establish limits of measurements which are subjected to the laws 
of quantum mechanics in a rigorous and unified manner.
In such a program, it is natural to expect that the uncertainty
principle will play a central role.  However, the
recent controversy \cite{Yue83,Cav85,88MS,89RS}, which arose in the
field of gravitational-wave detection, on the validity of 
the standard quantum limit for monitoring of the free-mass position 
\cite{Bra74,CTDSZ80} revealed a certain weakness
of our understanding of the Heisenberg uncertainty principle.  

Historically, Heisenberg established the uncertainty principle from
his analysis of {\em gedanken} experiments for certain processes of 
successive measurements \cite{Hei81}.  Thus his uncertainty principle 
is often stated \cite[p.~239]{vN55} in a form that a measurement of 
one variable from a conjugate 
pair disturbs the value of the conjugate variable no less than 
the order of $\hbar/$(accuracy of the measurement).
Nevertheless, we have not established a general theory of this kind 
of uncertainty principle, as pointed out by several authors
\cite{Bal70,MM90}.

Our uncertainty relation in current text books was first proved by
Robertson \cite{Rob29} by a simple mathematical reasoning with use 
of the Schwarz 
inequality.  However, it is often pointed out that Robertson's
uncertainty principle  does not mean the Heisenberg uncertainty principle.
Robertson's uncertainty principle is only concerned with state preparations
as in the statement that any state preparation gives an ensemble of objects
in which the product of the standard deviations of conjugate variables 
is greater than $\hbar/2$ \cite{Bal70,Hol82}.  

In this paper, we shall show how the Robertson uncertainty relation
gives certain intrinsic quantum limits of measurements in the most
general and rigorous mathematical treatment.  In Section 2, fragments
from our previous work on mathematical
foundations of quantum probability theory are given
(see, for the detail, \cite{83CR,84QC,85CA,85CC,86IQ,87MR}). 
In Section 3, some basic properties of root-mean-square error 
of measurement, called precision, introduced in \cite{88MS} is examined
and, in Section 4, a general lower bound of the product of precisions
arising in joint measurements of noncommuting observables is established.
This result is used to give a general proof of the uncertainty relation 
for the joint measurements which has been found by several authors 
\cite{ArtKel65,Hol72SQ,Yue82,ArtGoo88}.  In Section 5, 
we shall give a rigorous condition for holding of 
the standard quantum limit (SQL) for repeated measurements.  For this 
purpose, we shall examine another root-mean-square error, called 
resolution, introduced in \cite{88MS} and prove that if a measuring
instruments has no larger resolution than the precision then it obeys
the SQL.  As shown in \cite{88MS,89RS,90QP}, we can even construct many
linear models of position measurement which circumvent the above condition.
In Section 6, some conclusions from the present analysis will be
discussed.

  \section{Foundations of quantum probability}
  \label{2}\setcounter{equation}{0}
Let $\cH$ be a Hilbert space.  Denote by $\cL(\cH)$ the algebra of 
bounded linear operators on $\cH$,
by $\ta c(\cH)$  the space of trace class operators on $\cH$ and
by $\si c(\cH)$ the space of Hilbert-Schmidt class operators on $\cH$. 
A positive operator in $\ta c(\cH)$ with the unit trace is called 
a {\em density operator} and $\cS(\cH)$ stands for the space of
density operators on $\cH$.
Denote by $\cB(\R^d)$ the Borel $\si$-field of the Euclidean space
$\R^d$.
A map $X:\cB(\R^d) \to \cL(\cH)$ is called a 
{\em probability-operator-valued measure} (POM)
if it satisfies the following conditions (P1)--(P2):
\begin{list}{(P\arabic{bean})}{\usecounter{bean}}
\item For any sequence ${\langle}\De_i\mid i=1,2,\ldots{\rangle}$ of 
disjoint sets in $\cB(\R^d)$,
\[
X(\bigcup_{i=1}^{\infty}\De_i) = \sum_{i=1}^{\infty} X(\De_i),
\]
where the sum is convergent in the weak operator topology.
\item $X(\R) = 1$.
\end{list} 
A linear transformation $T:\ta c(\cH) \to \ta c(\cH)$ is called a {\em 
positive map\/} if $T(\rh) \ge 0$ for all $\rh \in 
\cS(\cH)$. We shall denote the space of all positive maps on $\ta c(\cH)$
by ${\rm P}(\ta c(\cH))$.  A map $\bX: \cB(\R^d) \to {\rm P}(\ta c(\cH))$ is 
called an {\em operation-valued measure} if it satisfies the following 
conditions (O1)--(O2):
\begin{list}{(O\arabic{bean})}{\usecounter{bean}}
\item For any sequence ${\langle}\De_i\mid i=1,2,\ldots{\rangle}$ of 
disjoint sets in $\cB(\R^d)$,
\[
     \bX(\bigcup _{i=1}^{\infty} \De_i) = \sum _{i=1}^{\infty} \bX(\De_i),
\]
\sloppy where the sum is convergent in the strong operator topology
of ${\rm P}(\ta c(\cH))$.
\item For any $\rho  \in  {\ta c}({\cal H})$,
\[ 
      \mbox{\rm Tr}[\bX({\bold R}^d )\rho ] = \mbox{\rm Tr}[\rho ].
\] 
\end{list}
An operation-valued measure $\bX: \cB(\R^d) \to {\rm P}(\ta c(\cH))$ is 
called a {\em completely positive operation-valued measure} (CPOM)
if it satisfies the following condition (O3):
\begin{list}{(O\arabic{bean})}{\usecounter{bean}}
\setcounter{bean}{2}
\item For any $\De \in \cB(\R^d)$, $\bX(\De)$ is a completely positive
map on $\ta c(\cH)$, i.e.,
\[
     \sum_{i,j=1}^{n} {\langle}\xi_i|\bX(\De)(|\et_i{\rangle}{\langle}\et_j|)|\xi_j{\rangle} \ge 0,
\]
for all $\De \in \cB(\R^d)$ and for all finite sequences 
$\xi_1,\ldots,\xi_n$ and $\et_1,\ldots,\et_n$ in $\cH$.
\end{list}
The {\em transpose} ${}^{t}T:{\cal L}({\cal H}) \to \cL({\cal H})$ 
of $T \in {\rm P}({\ta c}({\cal H}))$ is defined by the relation
                                                 \[
\mbox{\rm Tr}[{}^tT(a)\rho ] = \mbox{\rm Tr}[aT(\rho )],
                                                 \]\noindent
for all $a \in  {\cal L}({\cal H})$ and $\rho  \in  \ta c({\cal H})$. 
In this case, ${}^tT$ is also positive in the sense that $T(a) \ge 0$
for any $a \ge 0$ in ${\cal L}({\cal H})$. 
For any operation-valued measure $\bX$, the relation
                                                 \[
\hbX(\De) = {}^t\bX(\De)1 \qquad (\De \in \cB(\R^d)),
                                                 \]\noindent
determines a POM $\hbX$, called the POM {\em associated} with $\bX$.  
Conversely, 
any POM $X$ has at least one CPOM $\bX$ such that $X=\hbX$
\cite[Proposition 4.1]{84QC}.
POM's are called \lq\lq measurements'' in \cite{Hol82} and 
operation-valued measures are called \lq\lq instruments'' in 
\cite{DL70,Dav76}.  Our terminology is intended to be more neutral
in meanings in the physical context.

Suppose that a Hilbert space $\cH$ is the state space of
a quantum system $\bS$.  A {\em state} of $\bS$ is a density operator on $\cH$ and an 
{\em observable} of $\bS$ is a POM $A:\cB(\R) \to \cL(\cH)$ such
that $A(\De)$ is a projection for all $\De \in \cB(\R)$.
A state of the form $|\ps{\rangle}{\langle}\ps|$ for a unit vector $\ps \in \cH$
is called a {\em pure state} and $\ps$ is called a {\em vector state} of 
$\bS$.
A finite set $\{A_1,\ldots,A_n\}$ of observables is called 
{\em compatible} 
if $[A_i(\De_1),A_j(\De_2)] = 0$ for all $i,\ j = 1,\ldots,n$ and 
$\De_1,\ \De_2 \in \cB(\R)$.
The {\em joint probability distribution} of a compatible set 
$\{A_1,\ldots,A_n\}$ of observables in a state $\rh$,
denoted by $\Pr[A_1 \in \De_1,\ldots,A_n \in \De_n\|\rh]$, 
($\De_1,\ldots,\De_n \in \cB(\R)$), is given by the following 
{\em Born statistical formula}:
                                                 \[
\Pr[A_1 \in \De_1,\ldots,A_n \in \De_n\|\rh]
= \Tr[A_1(\De_1)\cdots A_n(\De_n)\rh].
                                                 \]\noindent
The symbol $\rh$ in the left-hand-side denotes the state for which the
probability distribution is determined.  For pure states 
$\rh = |\ps{\rangle}{\langle}\ps|$, the symbol $\ps$ will be sometimes used 
instead of $\rh$ in this and similar expressions,
and then we have
                                                 \[
\Pr[A_1 \in \De_1,\ldots,A_n \in \De_n\|\ps]
= {\langle}\ps|A_1(\De_1)\cdots A_n(\De_n)|\ps{\rangle}.
                                                 \]\noindent
By spectral theory, the relation
\[
\hA = \int_\R a A(da).
\]\noindent
sets up a one-to-one correspondence between observables $A$ and 
self-adjoint operators $\hA$.

In this paper, by a {\em measurement} we shall mean generally an experiment 
described as follows.  
Let $\bP$ be a quantum system, called a {\em probe system}, described by
a Hilbert space $\cK$.  The system $\bP$ is coupled to the system 
$\bS$ during a finite time interval from time $t$ to $t+\tilde{\tau}$.
Denote by $\hU$ the unitary operator on $\cH\otimes\cK$ corresponding
to the time evolution of the system $\bS+\bP$ from time $t$ to 
$t+\tilde{\tau}$. 
The time $t$ is called the {\em time of measurement} and the time 
$t+\tilde{\tau}$ is called the {\em time just after measurement}.
At the time just after the measurement, the systems $\bS$ and $\bP$ are
separated and in order to obtain the outcome of this experiment a 
compatible sequence ${\langle}M_1,\ldots,M_n{\rangle}$ of observables of
the system $\bP$ are measured by an ideal manner.  The observables
$M_1,\ldots,M_n$ are called the {\em meter observables}.  In order to assure
the reproducibility of this experiment, the probe system $\bP$ is always
prepared in a fixed state $\si$ at the time of measurement.
Thus the physical process of a given measurement is characterized by 
a 4-tuple $\cX = [\cK,\si,\hU,{\langle}\hM_1,\ldots,\hM_n{\rangle}]$, called a 
{\em measurement scheme}, consisting of a Hilbert
space $\cK$, a density operator $\si$ on $\cK$, a unitary 
operator $\hU$ on $\cH\otimes\cK$ and a compatible sequence
${\langle}\hM_1,\ldots,\hM_n{\rangle}$ of self-adjoint operators on $\cK$.
Every measurement scheme 
$\cX = [{\cal K},\sigma ,\hU,{\langle}\hM_1,\ldots,\hM_n{\rangle}]$ determines a
unique CPOM
$\bX:\cB(\R^n)\to {\rm P}(\ta c(\cH))$, called the CPOM of $\cX$, 
by the following relation
\beq{(cpom)}
\bX(\De_1\times\cdots\times\De_n)\rh 
= {\bf Tr}_{\cK}\left[\left(1\otimes M_1(\De_1)\cdots M_n(\De_n)\right)
\hU(\rh\otimes \si)\hU^{\dag}\right],
\eeq
for all $\rh \in \ta c(\cH)$ and $\De_1,\ldots,\De_n \in \cB(\R)$, where
${\bf Tr}_{\cK}$ stands for the partial trace operation of $\cK$.
Then the CPOM $\bX$ satisfies the following {\em Davies-Lewis postulates} 
(DL1)--(DL2) (cf.~\cite{DL70}):
\begin{list}{(DL\arabic{bean})}{\usecounter{bean}}
\item Measurement probability: If the state of the system $\bS$ at the 
time of measurement is $\rh$, then the probability distribution
$\Pr[X \in \De \|\rh]$, ($\De \in \cB(\R^n)$), of the outcome variable
$X$ of the measurement is   
                                                 \[
\Pr[X \in \De\|\rh] = \Tr[\bX(\De)\rh].
                                                 \]\noindent
\item State reduction: If the state of the system $\bS$ at the time
of measurement is $\rh$, then the measurement changes the state so that
the state $\rh_{\De}$, at the time just after measurement, of the 
subensemble $\bS_{\De}$ of the systems selected by the condition
$X \in \De$ is given by 
                                                 \[
\rh_{\De} = \frac{\bX(\De)\rh}{\Tr[\bX(\De)\rh]},
                                                 \]\noindent
for any $\De \in \cB(\R^n)$ with $\Pr[X \in \De\|\rh] \ne 0$.
\end{list}

Given an operation-valued measure $\bX:\cB(\R^n)\to {\rm P}(\ta c(\cH))$, 
any measurement scheme
$\cX = [{\cal K},\sigma ,\hU,{\langle}\hM_1,\ldots,\hM_n{\rangle}]$ which satisfies
Eq.~(\ref{(cpom)}) is called a {\em realization} of $\bX$.  
An operation-valued measure is called {\em realizable} if it has at least
one realization.  An importance of complete positivity for 
operation-valued measures is clear from the following.

\begin{Theorem}
An operation-valued measure $\bX:\cB(\R^n) \to {\rm P}(\ta c(\cH))$ is
realizable if and only if it is a CPOM.  In particular, every CPOM
$\bX:\cB(\R^n)\to {\rm P}(\ta c(\cH))$ has a realization $\cX 
= [{\cal K},\sigma ,\hU,{\langle}\hM_1,\ldots,\hM_n{\rangle}]$ such that $\si$ is a 
pure state and ${\rm dim}(\cH) = {\rm dim}(\cK)$.
\end{Theorem}

For a proof, see \cite[Section 5]{84QC}.  A consequence from the above 
theorem is the following version of the Naimark extension of POM's.

\begin{Corollary}
For any POM $X:\cB(\R^n) \to \cL(\cH)$, there exists a measurement
scheme $\cX=[\cK,|\ph{\rangle}{\langle}\ph|,\hU,{\langle}\hM_1,\ldots.\hM_n{\rangle}]$ satisfying
the relation
\bel{eq.2.2}
X(\De_1\times\cdots\times\De_n) 
= V^{\dag}[\hU^{\dag}(1\otimes M_1(\De_1)\cdots M_n(\De_n))\hU]V,
\eel{}
for all $\De_1,\ldots,\De_n \in \cB(\R^n)$, where $V$ is the isometry
from $\cH$ to $\cH\otimes\cK$ such that $V\ps = \ps\otimes\ph$
for all $\ps \in \cH$.
\end{Corollary}

A measurement scheme $\cX$ satisfying Eq.~(\ref{eq.2.2}) is called an
{\em interacting realization} of a POM $X$.  A definition of the 
non-interacting version of realizations of POM's appears in 
\cite[p. 68]{Hol82} and it should be noted that an interacting
realization determines the state reduction but 
a non-interacting one does not.

The outcome variable $X$ of a measurement scheme $\cX$ is generally 
called a {\em quantum random variable} (q.r.v.).  Thus any q.r.v. $X$ has a 
CPOM $\bX$ which determines the probability distributions of $X$. 
Let ${\langle}X_1,\ldots,X_n{\rangle}$ be a finite sequence of q.r.v.s of measurement
schemes $\cX_1,\ldots,\cX_n$
 and 
${\langle}\bX_1,\ldots,\bX_n{\rangle}$ the corresponding sequence of CPOM's.
Then, from (DL1) and (DL2), the joint probability distribution 
$\Pr[X_1\in \De_1,\ldots,X_n\in \De_n\|\rh]$, 
$(\De_1,\ldots,\De_n \in \cB(\R^n))$,
of ${\langle}X_1,\ldots,X_n{\rangle}$ in a state $\rh$ 
is given by the following {\em Davies-Lewis formula} \cite{DL70}:
\bel{}
\Pr[X_1\in \De_1,\ldots,X_n\in \De_n\|\rh]
= \Tr[\bX_n(\De_n)\cdots\bX_1(\De_1)\rh].
\eel
Let $\hH$ be the Hamiltonian of the system $\bS$ and $\hU_t
= e^{-it\hH/\hbar}$ the unitary operator of the time evolution.
For the Heisenberg system state $\rh$, we shall write,
                                                 \[
\rh(t) = \al(t)\rh = \hU_t\rh\hU_t^{\dag},
                                                 \]\noindent
for the time evolution of the states in the Schr\"{o}dinger picture.
Suppose that a finite sequence of measurements corresponding to a
sequence ${\langle}\bX_1,\ldots,\bX_n{\rangle}$ of CPOM is made successively
at time $(0<)t_1<\cdots<t_n$, where it is supposed that $\tilde{\ta}_i
\ll t_{i+1} - t_i$ for the durations $\tilde{\ta}_i$ of the coupling
of measurement $\bX_i$.   We shall denote by $X_i(t_i)$ the outcome
variable of the measurement $\bX_i$ at time $t_i$.  Then the 
joint probability distribution of the sequence 
${\langle}X_1(t_1),\ldots,X_n(t_n){\rangle}$ of the outcomes in the state
$\rh = \rh(0)$ is given by the following 
{\em Wigner-Davies-Lewis formula} \cite{Wig63,DL70}:
\begin{eqnarray*}
\lefteqn{
\Pr[X_1(t_1)\in\De_1,X_2(t_2)\in\De_2,\ldots,X_n(t_n)\in\De_n\|\rh]}\\
&=& \Tr[\bX_n(\De_n)\al(t_n-t_{n-1})\cdots
\bX_2(\De_2)\al(t_2-t_1)\bX_1(\De_1)\al(t_1)\rh], \\
& & \hfill (\De_1,\De_2,\ldots,\De_n\in\cB(\R)).
\end{eqnarray*}
 
Let $\bX: \cB(\R^n) \to \rP(\ta c(\cH))$ be a CPOM and $\rh$ a 
density operator on $\cH$.
A family $\{\rh_x \mid x \in \R^n\}$ of density operators on $\cH$
is called a family of {\em posterior states} for a {\em prior state} $\rh$ 
and a CPOM  $\bX$ if it satisfies the following conditions (PS1)--(PS2):
\begin{list}{(PS\arabic{bean})}{\usecounter{bean}}
\item The function $x \mapsto \rh_x$ is strongly Borel measurable.
\item For any $\De \in \cB(\R^n)$,
                                                 \[
\int_\De \rh_x\,\Tr[\bX(dx)\rh] = \bX(\De)\rh.
                                                 \]
\end{list}
A family of posterior states always exists for any prior state $\rh$
and it is unique in the following sense:  If
$\{\rho '_x\mid x \in  {\bold R} \}$ is another family of posterior states
for the prior state $\rho $, then $\rho'_x = \rho _x$ for
$\Tr[\bX(dx)\rh]$-almost everywhere \cite{85CA}.
Suppose that a measurement corresponding to $\bX$ is made for 
the system $\bS$ in a state $\rh$ at the time of measurement
and that the measurement gives the outcome $X=x$ $(x \in \R^n)$.
Let $\{\rh_x\mid x \in \R^n\}$ be a family of posterior states
for the prior state $\rh$.  Then, with probability 1, $\rh_x$ is 
the state of the system $\bS$ at the time just after the 
measurement.


  \section{Noise of approximate measurement}
  \label{3}\setcounter{equation}{0}
Let $\cH$ be a Hilbert space corresponding to a quantum system $\bS$.
Let $A$ be an observable of $\bS$. In this section, we consider a measuring 
instrument designed to measure 
the value of an observable $A$ and discuss the noise contained in 
outcomes from the measuring instrument.
Let $X$ be a q.r.v.\ representing the outcome from the measuring
instrument.  Then the probability distribution $\Pr[X \in \De\|\rh]$
of $X$ in a state $\rh$ of the system $\bS$ at the time of 
measurement is represented by a POM $\hbX$ for some CPOM $\bX$ 
satisfying
                                                 \[
\Pr[X \in \De\|\rh] = \Tr[\hbX(\De)\rh], \qquad (\De \in \cB(\R)). 
                                                 \]\noindent
For simplicity of notation, we shall write $X(\De) = \hbX(\De)$, 
$(\De \in \cB(\R))$.
Let $f$ be a real Borel function on $\R$.  The {\em expectation} 
$\Ex[f(X)\|\rh]$ of the q.r.v. $f(X)$ in a state $\rh$
is defined by
                                                 \[
\Ex[f(X)\|\rh] = \int_{\R} f(x) \Pr[X \in dx\|\rh],
                                                 \]\noindent
provided the integral is convergent.  Denote by $\widehat{f(X)}$
the symmetric operator defined by 
                                                 \belal{}\nonumber
{\displaystyle {\langle}\xi|\widehat{f(X)}|\xi{\rangle} 
= \int_\R f(x){\langle}\xi|X(dx)|\xi{\rangle}, 
\qquad (\xi \in \dom\left(\widehat{f(X)}\right)),}          \\
{\displaystyle  \dom\left(\widehat{f(X)}\right) 
= \{\xi\in\cH\mid \int_\R f(x)^2 {\langle}\xi|X(dx)|\xi{\rangle} < \infty \}.} 
                                                 \eelal
Then we have 
                                                 \[
\Ex[f(X)\|\ps] = {\langle}\ps|\widehat{f(X)}|\ps{\rangle},       
                                                 \]\noindent
for any vector state $\ps \in \dom(\widehat{f(X)})$.
The {\em variance} $\Var[X\|\rh]$ and the 
{\em standard deviation} $\De X[\rh]$ of $X$ in a state $\rh$ are
defined in the usual way, i.e.,
                                                     \belal{(3.a4)}
\Var[X\|\rh] = \Ex[X^2\|\rh] - \Ex[X\|\rh]^2,        \\
\De X[\rh] = \Var[X\|\rh]^{1/2}.                     
                                                     \eelal
We say that a POM $X$ has {\em finite second moment} in a state $\rh$
if $\Ex[X^2\|\rh] <\infty$, or equivalently, if $\De X[\rh] < \infty$.
Let $X$ be a POM with finite second moment in $\rh$.  Then we have
$\dom(\hX) \supset \ran(\sqrt{\rh})$ and that $\hX\sqrt{\rh}$ is 
a Hilbert-Schmidt operator \cite{Hol82}.  Thus, we shall write $\Tr[\hX^2\rh]
= \Tr[(\hX\sqrt{\rh})^{\dag}\hX\sqrt{\rh}]$. For any POM's $X$, $Y$ with 
$\De X[\rh]$, $\De Y[\rh]<\infty$, the expression 
$\Tr[\,[\hX,\hY]\rh]$ is defined by
\[
\Tr[\,[\hX,\hY]\rh] 
= \Tr[(\hX\sqrt{\rh})^{\dag}\hY\sqrt{\rh} 
    - (\hY\sqrt{\rh})^{\dag}\hX\sqrt{\rh}].
\]\noindent

By the Robertson uncertainty relation \cite{Rob29}, for any state 
$\rh$ and any pair of observables $A$, $B$ with $\De A[\rh]$, 
$\De B[\rh]<\infty$, we have
\bel{in.Rob}
\De A[\rh]\De B[\rh]\ge \frac{1}{2}\left|\Tr[\,[\hA,\hB]\rh]\right|.
\eel
The above relation is extended to any pair of POM's by Holevo
\cite[p.90]{Hol82},
i.e., for any pair of POM's $X$, $Y$ with $\De X[\rh]$,
$\De Y[\rh]<\infty$, we have
\bel{in.Hol}
\De X[\rh]\De Y[\rh]\ge \frac{1}{2}\left|\Tr[\,[\hX,\hY]\rh]\right|.
\eel

We say that a POM $X:\cB(\R) \to \cL(\cH)$ is {\em compatible} 
with an observable $A$ 
(or {\em $A$-compatible}, in short) if it satisfies the relation
                                                 \[
[X(\De_{1}),A(\De_{2})] = 0,
                                                 \]\noindent
for all $\De_{1}, \De_{2} \in \cB(\R)$.
Let $\rh$ be a state at the time of measurement.  For an $A$-compatible
POM $X$, the joint probability
distribution $\Pr[X \in \De_{1},A \in \De_{2}\|\rh]$ of $X$ and $A$
in a state $\rh$ is given by
                                                 \[
\Pr[X \in \De_{1},A \in \De_{2}\|\rh] 
= \Tr[X(\De_{1})A(\De_{2})\rh],
\qquad (\De_{1},\De_{2} \in \cB(\R)).
                                                 \]
By a notational convention, for a Borel measure $\mu$ on $\cB(\cR^2)$,
we shall write
\[
\iint_{\R^2} f(x,y)\nu(dx,dy) = \int_{\R^2} f(x,y)\mu(d(x,y)),
\]\noindent
where $\nu$ is the joint measure on $\cB(\R)\times\cB(\R)$ defined by
$\nu(\De_1,\De_2) = \mu(\De_1\times\De_2)$, ($\De_1, \De_2 \in \cB(\R)$),
and we shall write $\rL^p(\R^2,\nu(dx,dy)) = \rL^p(\R^2,\mu)$ for the
L$^p$-space of $\mu$. 
We define the {\em root-mean-square error} (or {\em precision}, in
short) $\ep[X|A,\rh]$ of $X$ for 
measurement of an observable $A$ in a state $\rh$ by the relation
\bel{(3.5)}
\ep[X\|A,\rh]^{2} =
\iint_{\R^{2}}\,(x - a)^{2}\,\Tr[X(dx)A(da)\rh].
\eel  
Obviously, $\ep[X|A,\rh]$ represents the root-mean-square deviation of the
outcome $X$ of the measurement from the outcome $A$ of the ideal measurement,
when these two were made simultaneously in the state $\rh$.

\begin{Lemma}\label{Lemma 3.1}
Let $\mu$ be a finite Borel measure
on ${\bold R}^2$.  Then the relation
\begin{equation}
  \iint _{{\bold R}^2} (x - y)^2\,\mu (dx\times dy) = 0
\label{(3.6)}
\end{equation}
holds if and only if for
any $\De_1$, $\De_2\in \cB(\R)$,
\begin{equation}
     \mu (\De_1\times \De_2) = \mu ((\De_1\cap \De_2)\times {\bold R}).
\label{(3.7)}
\end{equation}
\end{Lemma}

\begin{Proof} 
Suppose that Eq.~(\ref{(3.6)}) holds. Let 
$D = \{(x,y)\in {\bold R}^2| x = y\}$. Then
it follows from Eq.~(\ref{(3.6)}) that $\mu ({\bold R}^2\setminus D) = 0$. 
Then we have
\begin{eqnarray*}
    \mu (\De_1\times \De_2) 
&=& \mu ((\De_1\times \De_2)\cap D) = \mu ((\De_1\cap \De_2)\times {\bold R})\cap D) \\
&=& \mu ((\De_1\cap \De_2)\times {\bold R}).
\end{eqnarray*}
Conversely suppose
that Eq.~(\ref{(3.7)}) holds. Then we have
\begin{eqnarray*} 
    \mu (\De_1\times \De_2) 
&=& \mu ((\De_1\cap \De_2)\times {\bold R}) \\
&=& \int _{\De_1\cap \De_2} \mu (dx\times {\bold R}) \\
&=& \int _{\De_1}\delta _x(\De_2)\,\mu (dx\times {\bold R}) \\
&=& \int _{\bold R}\mu (dx\times {\bold R})
           \int _{\bold R}\chi _{\De_1\times \De_2}(x,y)\,\delta _x(dy),
\end{eqnarray*}
where $\delta _x$  is the Dirac measure of $x\in {\bold R}$. Thus we obtain
\begin{eqnarray*} 
  \iint _{{\bold R}^2} (x - y)^2\,\mu (dx\times dy)
&=& \int _{\bold R}\mu (dx\times {\bold R})
                           \int _{\bold R}(x - y)^2\,\delta _x(dy) \\
&=& 0.
\end{eqnarray*}
\end{Proof}

\begin{Theorem}\label{Theorem 3.2}
An $A$-compatible POM $X$ satisfies the relation
                                                              \bel{(3.8)}
\ep[X\|A,\rh] = 0,
                                                              \eel
for all density operator $\rh$ on $\cH$ if and only if 
$X = A$.
\end{Theorem}

\begin{Proof}
Obviously, if $X = A$ then Eq.~(\ref{(3.8)}) holds. Suppose that
Eq.~(\ref{(3.8)}) holds. Let $\rh \in \cS(\cH)$ 
and $\De\in \cB(\R)$.  Let $\mu$
be the Borel measure on ${\bold R}^2$ such that $\mu(\De_1\times \De_2) =
\Tr[X(\De_1)A(\De_2)\rh]$, ($\De_1, \De_2 \in \cB(\R)$). Then since 
$\ep[X\|A,\rh] = 0$,
we have $\mu (\De\times {\bold R})= \mu ({\bold R}\times \De)$ from Lemma 3.1, 
and hence
\[
     \mbox{\rm Tr}[A(\De)\rh ] 
= \mbox{\rm Tr}[A(\De)X({\bold R})\rh ] 
= \mbox{\rm Tr}[A({\bold R})X(\De)\rh ] = \mbox{\rm Tr}[X(\De)\rh ].
\]\noindent
Since $\rh $ and $\De$ are arbitrary, it is concluded that $A=X$.
\end{Proof}

\begin{Lemma}\label{Lemma 3.3}
For any $A$-compatible POM $X$, there exists a Hilbert space 
$\tilde{\cH}$, an isometry $V: \cH \to \tilde{\cH}$ and self-adjoint
operators $\tilde{X}$ and $\tilde{A}$ on $\tilde{\cH}$ satisfying
the following conditions:
\begin{list}{{\rm (\arabic{bean})}}{\usecounter{bean}}
\item $[\tilde{X},\tilde{A}] = 0$ and $\tilde{A}V = V\hA$.
\item For any $\rh \in \cS(\cH)$ and 
$f\in\rL^1\left(\R^2,\Tr[X(dx)A(da)\rh]\right)$,
                                                 \[
\iint_{\R^2} f(x,a)\,\Tr[X(dx)A(da)\rh] 
= \Tr[f(\tilde{X},\tilde{A})V\rh V^{\dag}].
                                                 \]\noindent
\end{list}
\end{Lemma}

\begin{Proof} 
Let ${M}: \cB(\R^2) \to \cL(\cH)$ be a POM such that 
${M}(\De_1\times\De_2)=X(\De_1)A(\De_2)$, for all 
$\De_1,\De_2\in\cB(\R)$.  Then, by the
Naimark extension of ${M}$, there exist a Hilbert space $\tilde{\cH}$,
an isometry $V: \cH \to \tilde{\cH}$ and a projection valued measure
$E: \cB(\R^2) \to \cL(\tilde{\cH})$ such that 
${M}(\De_1\times\De_2) = V^{\dag}E(\De_1\times\De_2)V$ for all 
$\De_1,\ \De_2 \in \cB(\R)$.  Let $\tilde{X}$ and $\tilde{A}$ be 
self-adjoint operators on $\tilde{\cH}$ defined by
                                            \begin{eqnarray*}
\lefteqn{\tilde{X} = \int_\R x \,E(dx\times\R),}  \\   
\lefteqn{\tilde{A} = \int_\R a \,E(\R\times da).}
                                            \end{eqnarray*}
Then the assertion follows from a straightforward verification.
\end{Proof}

An $A$-compatible POM $X$ is said to be {\em unbiased} if $\hA=\hX$;
in this case, we have. $\Ex[A\|\rh] = \Ex[X\|\rh]$ for all state
$\rh$ with $\De X[\rh]<\infty$.

\begin{Theorem}\label{Theorem 3.4}
Let $X$ be an unbiased $A$-compatible POM.  Then
for any state $\rh$ with $\De X[\rh]<\infty$, we have
                                                              \bel{(3.11)}
\ep[X\|A,\rh]^2 = \De X[\rh]^2 - \De A[\rh]^2.
                                                              \eel
\end{Theorem}

\begin{Proof}
By Lemma~\ref{Lemma 3.3}, we have
\begin{eqnarray*}
\ep[X\|A,\rh]^2 
&=& \iint_{\R^{2}}\,(x - a)^{2}\,\Tr[X(dx)A(da)\rh]  \\
&=& \Tr[(\tilde{X}-\tilde{A})^{2}V\rh V^{\dag}] \\
&=& \Tr[\tilde{X}^{2}V\rh V^{\dag}] - \Tr[\tilde{A}^{2}V\rh V^{\dag}] \\
&=& \De X[\rh]^2 - \De A[\rh]^2.
\end{eqnarray*}
\end{Proof}

\begin{Remark}
When $X$ is $A$-compatible but $\hA \ne \hX$, we have
\[
\ep[X|A,\rh]^2 = \De X[\rh]^2 - \De \hX[\rh]^2 + \Tr[(\hA - \hX)^2\rh],
\]\noindent
where $\De \hX[\rh]^2 = \Tr[\hX^2\rh] - \Tr[\hX\rh]^2$.
It follows that $\ep[X|A,\rh]$ has a lower bound such that
$\ep[X|A,\rh]^2 \ge \Tr[(\hA - \hX)^2\rh]$.
\end{Remark}

  \section{Uncertainty principle for joint measurements}
  \label{4}\setcounter{equation}{0}

Consider a measuring instrument with two output variables $X$, $Y$
designed to measure the values of observables $A$, $B$ of a quantum
system $\bS$ described by a Hilbert space $\cH$.  Let $M:\cB(\R^2)
\to \cL(\cH)$ be the joint POM of the pair ${\langle}X,Y{\rangle}$, and $\rh$ be 
a state of $\bS$ at the time of measurement.  Then we have
\[
\Pr[X \in \De_1,Y \in \De_2\|\rh] = \Tr[M(\De_1\times\De_2)\rh],
\]\noindent
for all $\De_1$, $\De_2 \in \cB(\R)$.  Let ${\langle}M_X,M_Y{\rangle}$ be 
the pair of marginal POM's of $M$, i.e., $M_X(\De) = M(\De\times\R)$,
$M_Y(\De) = M(\R\times\De)$, ($\De \in \cB(\R)$).  Then $M_X$ and 
$M_Y$ are the POM's of q.r.v.s $X$ 
and $Y$, respectively, and hence it is natural to assume that $M_X$
is an unbiased $A$-compatible POM and that $M_Y$ is an unbiased 
$B$-compatible POM.  In this case, it is known 
\cite{ArtKel65,Hol72SQ,Yue82,ArtGoo88} that $\De X[\rh]$
and $\De Y[\rh]$ obeys a more stringent uncertainty relation than
the Robertson-Holevo relation (\ref{in.Hol}).  A general proof
of this fact is given below along with the ideas in \cite{ArtGoo88}.

A pair ${\langle}X,Y{\rangle}$ of POM's is called a {\em coexistent} pair 
if there is a POM $M:\cB(\R^2) \to \cL(\cH)$ such that
$X(\De) = M(\De\times\R)$, $Y(\De) = M(\R\times\De)$, 
for all $\De \in \cB(\R)$.

\begin{Theorem}
Let $\hA$, $\hB$ be self-adjoint operators on a Hilbert space 
$\cH$.  Let ${\langle}X,Y{\rangle}$ be a coexistent pair of POM's such
that $X$ is an unbiased $A$-compatible POM and $Y$ is an unbiased
$B$-compatible POM.  Then, for any state $\rh$ with 
$\De X[\rh], \De Y[\rh]<\infty$, we have
\[
{\rm (1)}\  \ep[X\|A,\rh]\ep[Y\|B,\rh] \ge \frac{1}{2}|\Tr[\,[\hA,\hB]\rh]|,
\]
\[
{\rm (2)}\  \De X[\rh]\De Y[\rh] \ge \left|\Tr[\,[\hA,\hB]\rh]\right|.
\]
\end{Theorem}

\begin{Proof} For simplicity, we shall prove the case where
$\rh=|\ps{\rangle}{\langle}\ps|$.
Let $M:\cB(\R^2) \to \cL(\cH)$ be a POM such that $M(\De\times\R)
= X(\De)$ and $M(\R\times\De) = Y(\De)$.  Let 
$\cX = [\cK,|\ph{\rangle}{\langle}\ph|,\hU,{\langle}\hM_1,\hM_2{\rangle}]$ be an interacting 
realization of
$M$.  Set $X_1=X$, $X_2=Y$, $A_1=A$, and $A_2=B$.
Define noise operators $\hN_i$ $(i=1,2)$ by the relation
\[
\hN_i = \hU^{\dag}(1\otimes \hM_i)\hU - \hA_i\otimes 1.
\]\noindent
Then we have
                                                 \begin{eqnarray*}
{\langle}\ps\otimes\ph|\hN_i|\ps\otimes\ph{\rangle}
&=& {\langle}\ps\otimes\ph|\hU^{\dag}(1\otimes \hM_i)\hU|\ps\otimes\ph{\rangle} 
- {\langle}\ps|\hA_i|\ps{\rangle}                                \\
&=& {\langle}\ps|\hX_i|\ps{\rangle} - {\langle}\ps|\hA_i|\ps{\rangle}              \\
&=& 0,
                                                 \end{eqnarray*}
and hence
                                                 \begin{eqnarray*}
\De \hN_i[\ps\otimes\ph]^2
&=& {\langle}\ps\otimes\ph|\hN_{i}^2|\ps\otimes\ph{\rangle}    \\
&=& {\langle}\ps\otimes\ph|(\hU^{\dag}(1\otimes M_i)\hU - \hA_i\otimes 1)^2
|\ps\otimes\ph{\rangle}\\ 
&=& {\langle}\ps|\hX_i^2|\ps{\rangle} - {\langle}\ps|\hA_i^2|\ps{\rangle}          \\
&=& \ep[X_i|A_i,\ps]^2.
                                                 \end{eqnarray*}
On the other hand, from the relations ($i, j = 1, 2$)
                                                 \begin{eqnarray*}
{\langle}\ps\otimes\ph|\hU^{\dag}(1\otimes M_i)\hU(\hA_j\otimes 1)|\ps\otimes\ph{\rangle}
&=& {\langle}\ps\otimes\ph|\hU^{\dag}(1\otimes M_i)\hU|(\hA_j\ps)\otimes\ph{\rangle} \\
&=& {\langle}\ps|\hA_i\hA_j|\ps{\rangle},
                                                 \end{eqnarray*}
we have
                                                 \[
{\langle}\ps\otimes\ph|[\hN_{1},\hN_{2}]|\ps\otimes\ph{\rangle}   
= {\langle}\ps|[\hA,\hB]|\ps{\rangle}.
                                                 \]\noindent 
Thus by the Robertson uncertainty relation we have
                                                 \begin{eqnarray*}
\ep[X|A,\ps]\ep[Y|B,\ps] 
&=& \De \hN_1[\ps\otimes\ph]\De \hN_2[\ps\otimes\ph] \\
&\ge& \frac{1}{2}\,|{\langle}\ps\otimes\ph|[\hN_{1},\hN_{2}]|\ps\otimes\ph{\rangle}|\\
&=&   \frac{1}{2}\,|{\langle}\ps|[\hA,\hB]|\ps{\rangle}|.
                                                 \end{eqnarray*}
This concludes (1).  From this relation, Theorem~\ref{Theorem 3.4} and
the Robertson uncertainty relation, we obtain
                                                 \begin{eqnarray*}
\De X[\rh]^2\De Y[\rh]^2
&=&   (\ep[X|A,\rh]^2+\De A[\rh]^2)(\ep[Y|B,\rh]^2+\De B[\rh]^2) \\
&\ge& (\ep[X|A,\rh]\ep[Y|B,\rh]+\De A[\rh]\De B[\rh])^2          \\
&\ge& |\Tr[\,[\hA,\hB]\rh]|^2.      
                                                 \end{eqnarray*}
This proves (2).
\end{Proof}

  \section{Standard quantum limit for repeated measurements}
  \label{5}\setcounter{equation}{0}
Let $\bX:\cB(\R) \to \rP(\ta c(\cH))$ be a CPOM and $A$ an observable of 
a system $\bS$ corresponding to $\cH$.
We define the {\em root-mean-square scatter} (or {\em resolution}, in short)
$\si[\bX\|A,\rh]$ of a CPOM $\bX$ for measurement of an observable $A$ 
in a state $\rh$ by the relation  
                                                 \bel{eq.5.1}
\si[\bX\|A,\rh]^2 = \iint_{\R^2} (x - a)^2 \,\Tr[A(da)\bX(dx)\rh].
                                                 \eel
Let $\{\rh_x\mid x \in \R\}$ be a family of posterior state
for a prior state $\rh$ and $\bX$.  Then, we have
                                                 \bel{resolution}
\si[\bX\|A,\rh]^2 
= \int_{\R} \Tr[\bX(dx)\rh] \int_{\R}(x - a)^2 \,\Tr[A(da)\rh_x].
                                                 \eel

         \begin{Theorem}\label{Theorem 5.1}
Let $\hA$ be a self-adjoint operator on a Hilbert space $\cH$.  Let
$\bX:\cB(\R) \to \rP(\ta c(\cH))$ be a CPOM, $\rh$ a density
operator on $\cH$ with $\De A[\bX(\R)\rh]$, $\De \hbX[\rh] <\infty$ and
$\{\rh_x\mid x \in \R\}$ a family of posterior states for $\rh$ and $\bX$.
Then we have 
                                                 \[
\si[\bX\|A,\rh]^2 
= \int_\R \De A[\rh_x]^2\,\Tr[\bX(dx)\rh]
+ \int_\R (\Tr[\hA\rh_x] - x)^2\,\Tr[\bX(dx)\rh].
                                                 \]
         \end{Theorem}

         \begin{Proof}
From $\De A[\bX(\R)\rh]<\infty$, we obtain
                                                 \[
\int_\R \Tr[\hA^2\rh_x]\,\Tr[\bX(dx)\rh]
= \Tr[\hA^2\bX(\R)\rh] < \infty,
                                                 \]\noindent
and hence $\Tr[\hA^2\rh_x] <\infty$, $\Tr[\bX(dx)\rh]$-almost everywhere.
Thus the assertion follows from Eq.~(\ref{resolution}) 
and the relations
                                                 \begin{eqnarray*}
\int_\R (a - x)^2\,\Tr[A(da)\rh_x] 
&=& \Tr[\hA^2\rh_x] - 2x\Tr[\hA\rh_x] + x^2                   \\
&=& \Tr[\hA^2\rh_x] - \Tr[\hA\rh_x]^2 + (\Tr[\hA\rh_x] - x)^2 \\
&=& \De A[\rh_x]^2                    + (\Tr[\hA\rh_x] - x)^2.
                                                 \end{eqnarray*}
         \end{Proof}

Let $\bX$ be a CPOM of a measuring instrument with one output variable $X$
designed to make an unbiased measurement of an observable $A$ 
of a system $\bS$ corresponding to a Hilbert space $\cH$.  
Suppose that the system $\bS$ undergoes unitary evolution during
the time $\ta$ between two identical measurements described by the CPOM $\bX$.
Let $\hU_\ta $ be the unitary operator of the time evolution of
the system $\bS$, i.e., $\hU_{\ta}=e^{-i\ta\hH/\hbar}$, where $\hH$ is the
Hamiltonian of $\bS$.
Suppose that the system $\bS$ is in a state $\rh$ at the time
of the first measurement. Then at the time just after the first 
measurement (say, $t = 0$) the system is in a posterior state $\rh_x$
with the probability distribution $\Pr[X\in dx\|\rh] = \Tr[\bX(dx)\rh]$. 
From this outcome $X=x$,
the observer makes a prediction $X(\tau)=h(x)$ for the outcome
of the second measurement at $t= \ta$.  Then the squared uncertainty
of this prediction is
                                                 \begin{eqnarray}
 \De[\ta,\rh,x]^2
&=& \int_\R (a - h(x))^2 \Pr[X \in da\|\rh_x(\ta)]    \nonumber\\
&=& \int_\R (a - h(x))^2 \,\Tr[\bX(da)\al(\ta)\rh_x]. \label{eq.5.3}
                                                 \end{eqnarray}
As to determination of $h(x)$, the following mean-value-prediction strategy
is naturally adopted:
                                                 \begin{equation}
h(x) = \Tr[\rh_{x}\hA(\ta)],                     \label{eq.5.4}
                                                 \end{equation}
where
                                                 \begin{equation}
 \hA(\ta) = \hU_{\ta}^{\dag}\hA(0)\hU_{\ta}.     \label{eq.5.5}
                                                 \end{equation}
The predictive uncertainty $\De[\ta,\rh]$ of this repeated
measurement with the prior state $\rh$ and the time duration $\ta$ is defined
as the root-mean square of $\De[\ta,\rh,x]$ over all outcomes $X=x$ of the
first measurement, i.e.,
                                                 \begin{eqnarray}
 \De[\ta,\rh]^2 
&=& \int \De[\ta,\rh,x]^2 \Pr[X\in dx\|\rh]      \nonumber\\
&=& \iint_{\R^2} (a - h(x))^2 \,\Tr[\bX(da)\al(\ta)\bX(dx)\rh].
                                                 \label{eq.5.6}
                                                 \end{eqnarray}
 \begin{Theorem}\label{41T2}
Let $\bX:\cB(\R) \to \rP(\ta c(\cH))$ be an unbiased $A$-compatible CPOM
and $\rh$ a density operator on $\cH$ with $\De A(0)[\bX(\R)\rh]$,
$\De A(\ta)[\bX(\R)\rh] <\infty$.
If the relation
                                                 \begin{equation}
 \si[\bX\|A,\rh] \le \ep[\hbX\|A,\al(\ta)\bX(\R)\rh],
                                                 \label{res.le.pre}
                                                 \end{equation}
holds then we have
                                                 \begin{equation}
 \De[\ta,\rh]^2 
\ge \left|\Tr[\,[\hA(0),\hA(\ta)]\bX(\R)\rh]\right|.
                                                 \label{in.sql}
                                                 \end{equation}
 \end{Theorem}

\begin{Proof}
From Eqs.~(\ref{eq.5.3})--(\ref{eq.5.5}) and Theorem~\ref{Theorem 3.4},
                                                 \begin{eqnarray*}
\De[\ta,\rh,x]^2 
&=& \De X[\al(\ta)\rh_x]^2                       \\
&=& \ep[X\|\al(\ta)\rh_x]^2 + \De A(\ta)[\rh_x]^2,
                                                 \end{eqnarray*}
and hence by Theorem~\ref{Theorem 5.1} and the Robertson uncertainty
principle,
                                                 \begin{eqnarray*}
 \De[\ta,\rh]^2
&=& \int_{\R} \ep[\bX\|\al(\ta)\rh_x]^2 
                  + \De A(\ta)[\rh_x]^2\,\Tr[\bX(dx)\rh] \\
&=&   \ep[\hbX\|A,\al(\ta)\bX(\R)\rh]^2
          + \int_\R \De A(\ta)[\rh_x]^2\,\Tr[\bX(dx)\rh] \\
&\ge & \si[\bX\|A,\rh]^2 
          + \int_\R \De A(\ta)[\rh_x]^2\,\Tr[\bX(dx)\rh] \\
&\ge & \int_\R \De A(0)[\rh_x]^2 + \De A(\ta)[\rh_x]^2\,\Tr[\bX(dx)\rh] \\
&\ge & \int_\R 2\De A(0)[\rh_x]\De A(\ta)[\rh_x]\,\Tr[\bX(dx)\rh] \\
&\ge & \int_\R 
\left|\Tr[\,[\hA(0),\hA(\ta)]\rh_x]\right|\,\Tr[\bX(dx)\rh] \\
&\ge&   \left|\Tr[\,[\hA(0),\hA(\ta)]\bX(\R)\rh]\right|.
                                                  \end{eqnarray*}
\end{Proof}

The bound (\ref{in.sql}) is called the 
{\em standard quantum limit} (SQL) for repeated measurements with interval 
$\ta$
of an observable $A$.  For the case where $A$ is the position observable
$x$ of a free-mass $m$, relation (\ref{in.sql}) is reduced to the
relation
\bel{in.SQL}
\De[\ta,\rh]^2 \ge \frac{\hbar\ta}{m},
\eel
which was posed in \cite{Bra74,CTDSZ80}
and the validity of this standard quantum limit was the subject of a long
controversy \cite{Yue83,Cav85,Ni86,88MS}.
By the above theorem, any measuring instrument which beats the SQL must
have the resolution lager than the precision. In 
\cite{Yue83}, Yuen pointed out a flaw in the original derivation of 
the SQL (\ref{in.SQL}) and proposed an idea of using contractive states to 
beat the SQL.  A model which clears the above condition 
and beats the SQL
was successfully constructed in our previous work \cite{88MS,89RS} as
a realization of Gordon-Louisell measurement $\{|\mu\nu a\om{\rangle}{\langle}a|\}$
\cite{GL66}, where $|\mu\nu a\om{\rangle}$ is a contractive state and 
$|a{\rangle}$ is a position eigenstate.  Many linear-coupling models of position 
measurements which violates condition (\ref{res.le.pre}) are constructed in 
\cite{90QP}. 

  \section{Concluding remarks}
  \label{6}\setcounter{equation}{0}

We have discussed quantum mechanical limitations on joint measurements
and repeated measurements of a single object. It is shown that the 
uncertainty principle for joint measurements of noncommuting observables 
holds generally with a more stringent limit than the one usually supposed by 
the Robertson uncertainty relation.  On the other hand, the SQL, which is
also usually supposed from the Robertson uncertain relation, for repeated
measurements of a single observable does not generally hold unless a certain
additional condition is satisfied.  The difference between these two problem
is clear from the difference between those two uncertainties defined by 
Eq.~(\ref{(3.5)}) and Eq.~(\ref{eq.5.6}) for which the optimizations are
required.  The crucial point is that in the latter problem we can use
the result of the first measurement to predict the result of the second
and hence the prediction can be based on posterior probability.  However, 
in the problem of joint measurements, we are required to predict two outcomes
only from prior probability given by the prior state.  Thus we can 
circumvent the uncertainty principle in the problem of repeated measurements,
when the measurement changes the prior state to the posterior state which
has deterministic information about the {\em future value} of the 
observable to be measured.  
Of course, this future value must be significantly uncertain,
if the prior state is of deterministic information about the {\em
present value} and the measurement is {\em not} carried out.  However, 
some measurement can give this present value precisely and further
leaves the object in a state with deterministic information about 
the future value.  Thus monitoring a mass in this way can give a precise
information about {\em classical} force which drives the mass. 

\bigskip

The author wishes to thank Professor Horace P. Yuen for hospitality at 
Northwestern University and Professor Roy J. Glauber for hospitality 
at Harvard University during his leave in 1988--1990.  This work is
supported in part by Hamamatsu Photonics K. K.
 


  \end{document}